\documentclass[12pt]{amsart}
\usepackage{amsmath,amsfonts,amssymb} \usepackage{color}
\usepackage[all,cmtip]{xy}
\usepackage{graphicx}

\newcommand{\corr}[1]{\langle {#1} \rangle}

\newcommand{\bt}{{\bf t}} \newcommand{\bT}{{\bf T}}

  \newcommand{\cO}{\mathcal{O}}

 \newcommand{\cV}{\mathcal{V}} 
 \newcommand{\bZ}{\mathbb{Z}}
 
 \newcommand{\pd}{\partial}
\newcommand{\Mbar}{\overline{\mathcal M}}

  \DeclareMathOperator{\res}{res}

\newcommand{\be}{\begin{equation}}
\newcommand{\ee}{\end{equation}}
\newcommand{\bea}{\begin{eqnarray}}
\newcommand{\eea}{\end{eqnarray}}
\newcommand{\ben}{\begin{eqnarray*}}
\newcommand{\een}{\end{eqnarray*}}

\newcommand{\half}{\frac{1}{2}}

\newtheorem{cor}{Corollary}[section]

 \newtheorem{thm}[cor]{Theorem}
\theoremstyle{remark}

\definecolor{A}{rgb}{.75,1,.75}

\definecolor{yellow}{rgb}{1,1,0}
\definecolor{orange}{rgb}{1,.7,0}
\definecolor{red}{rgb}{1,0,0}
\definecolor{white}{rgb}{1,1,1}

 \makeindex
\begin{document}
\title
{Solution of W-Constraints for R-Spin Intersection Numbers}

\author{Jian Zhou}
\address{Department of Mathematical Sciences\\Tsinghua University\\Beijng, 100084, China}
\email{jzhou@math.tsinghua.edu.cn}

\begin{abstract}
We present a solution to the W-constraints satisfied by the intersection numbers on the moduli spaces
of r-spin curves.
We make use of a grading suggested by the selection rule for the correlators 
determined by the geometry of the moduli space. 
\end{abstract}

\maketitle

\section{Introduction}

The famous Witten Conjecture \cite{Witten1} proved by Kontsevich \cite{Kon} relates intersection numbers on moduli spaces of stable curves
to the KdV hierarchy.
Witten \cite{Witten2} introduce r-spin curves, their moduli spaces and
conjectured that the intersection numbers on them are related to generalized KdV hierarchies (Gelf'and-Dickey hierarchies).
This conjecture has been proved by Faber-Shadrin-Zvokine \cite{Fab-Sha-Zvo}.
More recently,
such problems have been also been studied from the point of view of Givental quantization formalism \cite{Giv, Giv-Mil, Fre-Giv-Mil}
and also from the point of view of Fan-Jarvis-Ruan-Witten theory \cite{FJRW}.

As pointed out by Witten \cite{Witten1},
together with the string equation,
the KdV hierarchy completely determines all the intersection numbers of psi-classes on $\Mbar_{g,n}$;
similarly,
the generalized KdV hierarchy together with the string equation
completely determine the r-spin intersection numbers \cite{Witten2}.
Liu, Vakil and Xu \cite{Liu-Vakil-Xu} have developed an algorithm
to compute the r-spin intersection numbers based on such these facts.
Some explicit results can be found in {\em loc. cit.} for $r=5$ and similar results for $r=4$ and $5$ can
be found in an earlier paper by Liu and Xu \cite{Liu-Xu}.
Such results match with results by other methods obtained by e.g. Shadrin \cite{Shadrin}, Brezin and Hikami \cite{Brezin-Hikami}.

It is well-known that such intersection numbers also satisfy linear constraints called the Virasoro constraints in the $r=2$ case
and the $W$-constraints in the general case.
For derivations of the equivalence of such linear constraints with the generalized KdV hierarchy together with the string equation,
see Dijkgraaf-Verlinde-Verlinde \cite{DVV}, Fukuma-Kawai-Nakayama \cite{Fukuma-Kawai-Nakayama1, Fukuma-Kawai-Nakayama2},  Goeree \cite{Goeree}, and Kac-Schwarz \cite{Kac-Schwarz}.
For algebraic background on $W$-algebras,
we refer to Fateev-Lukyanov \cite{Fateev-Lukyanov}, Feigin-Frenkel \cite{Feigin-Frenkel1, Feigin-Frenkel2}
and Frenkel-Kac-Radul-Wang \cite{FKRW}.
More recently,
Bakalov and Milanov \cite{Bakalov-Milanov1, Bakalov-Milanov2}
constructed W-constraints for Frobenius manifolds associated with simple singularities,
and they conjectured their constraints uniquely determine the partition function up to a factor.

Recently, Liu, Yang and Zhang \cite{Liu-Yang-Zhang} prove the conjecture of Bakalov and Milanov and extend their construction
of to more general Frobenius manifolds.
A key ingredient of this proof is a natural grading of the coupling constants.
We will use a different grading which is compatible with selection rule
for nonvanishing correlator as determined by the geometry of the moduli spaces.
Then we obtain a solution of the W-constraints similar
to the case of Witten-Kontsevich partition function as in Alexandrov \cite{Alexandrov}.

We will focus on the type A case,
i.e., the r-spin curve intersection numbers, in this paper.
The method also works in the D and E cases,
and the details will be presented in a separate paper.

The rest of the paper is arranged as follows.
In Section 2 we recall some well-known backgrounds on KP hierarchy and its reductions into generalized KdV hierarchies.
In Section 3 we recall the partition of Witten's r-spin intersection number
and the generalized KdV hierarchies satisfied by them.
We specify the change of variables to make Witten's original formulation compatible with the standard notations.
The W-constraints for r-spin numbers will be given in Section 4
and their solutions will be presented in Section 5.
In Section 6 we present some examples that verify our solution.

\section{KP Hierarchy and Its Reductions}

For reference,
see Miwa-Jimbo-Date \cite{MJD}.

\subsection{The KP hierarchy}

Consider the algebra of pseudodifferential operators of the form
$$P = \sum_{j \leq n} f_j(x) D^j$$
for some $n \in \bZ$,
where $D =\frac{\pd}{\pd x}$.
One can define the multiplications of such operators using:
\be
D^n \circ f = \sum_{j \geq 0} \binom{n}{j} \cdot (D^j f) \cdot D^{n+j}.
\ee
Define
\bea
&& P_+ = \sum_{j \geq 0} f_j(x) D^j, \\
&& P_- = \sum_{j < 0} f_j(x) D^j, \\
&& P^* =  \sum_{j \leq n} f_j(x) (-D)^j, \\
&& \res (P) = f_{-1}(x).
\eea

The KP hierarchy is a system of evolution equations on the space of pseusodifferential operators of the form:
\be
Q = D + \sum_{j=1}^\infty f_j(\bT) D^{-j},
\ee
given by
\be
\frac{\pd}{\pd T_n} Q = [(Q^n)_+, Q] = [Q, (Q^n)_-].
\ee
Here $\bT = T_1, T_2, \dots$ and $T_1 = x$.

\subsection{The dressing operator and the wave functions}

Write $Q$ as $Q = WDW^{-1}$,
where
\be
W = 1+ \sum_{j=1}^\infty w_j D^{-j}.
\ee
If $W$ satisfies
\be \label{eqn:Wtn}
\frac{\pd}{\pd T_n} W = - (Q^n)_- W,
\ee
then the KP hierarchy is satisfied by $Q$.
The wave function of the KP hierarchy is defined by:
\be
w(\bT, z) = W e^{\xi(\bT; z)} = (1 + \sum_{j=1}^\infty w_j(\bT) z^{-j}) \cdot e^{\xi(\bt;z)},
\ee
where $\xi(\bT; z) = \sum_{j=1}^\infty T_j z^j$, $t_1 = x$.
Applying $Q$ to $w$ one gets:
\be
Q w(\bT;z) = z \cdot w(\bT;z).
\ee
Applying $\frac{\pd}{\pd T_n}$ to $w$ one gets:
\be
\frac{\pd}{\pd T_n} w(\bT; z) = (Q^n)_+w(\bT; z).
\ee

Similarly,
the adjoint wave function $w^*(\bT;z)$ is defined by
\be
w^*(\bT; z) = (W^{-1})^*e^{-\xi(\bT;z)}.
\ee

\subsection{Tau-function and the vertex operators}

It turns out that there is a function $\tau(\bt)$ such that
\bea
&& w(\bT; z)
= \frac{\tau\big( T_1-  \frac{1}{z}, T_2- \frac{1}{2z^2}, \dots\big)}{\tau(\bT)} e^{\xi(\bT;z)}, \\
&& w^*(\bt; z)
= \frac{\tau\big( T_1 +  \frac{1}{z}, tT_2 + \frac{1}{2z^2}, \dots\big)}{\tau(\bT)} e^{-\xi(\bT;z)}.
\eea
In terms of the vertex operators
\be
X(\bT;z) = e^{\xi(\bt; z)} e^{-\xi(\tilde{D}, 1/z)}, \quad
\tilde{X}(\bT; z) = e^{-\xi(\bt; z)} e^{\xi(\tilde{D}, 1/z)},
\ee
where $\tilde{D} = (\frac{\pd}{\pd t_1}, \frac{1}{2} \frac{\pd}{\pd t_2}, \dots)$,
one has
\be
w(\bT;z) = \frac{X(\bT;z)\tau(\bT)}{\tau(\bT)}, \quad
w^*(\bT;z) = \frac{\tilde{X}(\bT;z) \tau(\bT)}{\tau(\bT)}.
\ee

The vertex operators can be rewritten in terms of the following field of operators:
\be
\phi(\bT;z) = \sum_{n \in \bZ} \phi_n(\bT) z^{-n},
\ee
where $\phi_n$'s are operators defined by:
\be
\phi_n(\bT) = \begin{cases}
- \frac{1}{n} \frac{\pd}{\pd T_n}, & n > 0, \\
T_{-n} \cdot, & n < 0, \\
0, & \text{otherwise}.
\end{cases}
\ee
For $n > 0$,
the operators $\phi_{-n}$'s are creators,
and the operators $\phi_n$'s are annihilators.
In other words,
\be
\phi(\bT; z) = \sum_{n=1}^\infty z^{-n} T_n \cdot
- \sum_{n=1}^\infty z^n \frac{1}{n} \frac{\pd}{\pd T_n}.
\ee
Then
\be
X(\bT; z) = :e^{\phi(z)}:, \quad
\tilde{X}(\bT;z) = :e^{-\phi(\bT;z)}:.
\ee
Here $:\cdot:$ means the normal ordering,
i.e.,
the annihilators are always put on the right of the creators.

\subsection{The $r$-th reduction of KP hierarchy}

For a positive integer $r$,
the $r$-th generalized KdV hierarchy (also called the Gelf'and-Dikii hierarchy)
is obtained from the KP hierarchy by imposing the condition that the tau-function
$\tau(\bT)$ does not depend on $t_k$ when $k \equiv 0 \pmod{r}$.
By \eqref{eqn:Wtn},
$Q^r_-=0$,
hence one can write $Q^r$ as some operator
\be
L = D^r + u_{r-2} D^{r-2} + \cdots + u_1 D + u_0
\ee
for some functions $u_0, \dots, u_{r-2}$.
One can then rewrite the $r$-th reduced KP hierarchy in terms of the operator $L$ as follows:
\be
\frac{\pd}{\pd t_n} L = [(L^{n/r})_+, L] = [L, (L^{n/r})_-].
\ee

\section{Partition Function of Witten's R-Spin Intersection Numbers}

\subsection{Witten's r-spin intersection numbers}
They are defined by Witten \cite{Witten2} as follows:
\be
\corr{ \tau_{m_1, a_1} \cdots \tau_{m_n, a_n} }_g
= \frac{1}{r^g} \int_{\Mbar^{1/r}_{g,n}} \prod_{i=1}^n \psi(x_i)^{m_i} \cdot e(\cV).
\ee
These are nonzero only when the following selection rule is satisfied:
\be \label{eqn:SelRule}
(r+1)(2g-2) + r n = r \sum_{i=1}^n m_i + \sum_{i=1}^n a_i,
\ee
and
\be
a_i \neq r-1, \quad i = 1, \dots, n.
\ee
These intersection numbers satisfy the string equation
\be \label{eqn:String}
\corr{\tau_{0,0} \prod_{i=1}^n \tau_{m_i, a_i}}_g
= \sum_{j=1}^n \corr{\tau_{m_j-1, a_j} \cdot \prod_{\substack{1 \leq i \leq n \\ i \neq j}}
\tau_{m_i, a_i}}_g,
\ee
and the dilaton equation:
\be \label{eqn:Dilaton}
\corr{\tau_{1,0} \prod_{i=1}^n \tau_{m_i, a_i}}_g
= (2g-2+n) \cdot \corr{ \prod_{i=1}^n \tau_{m_i, a_i}}_g.
\ee
In genus $0$,
the following was calculated in \cite{Witten2}:
\be
\corr{\tau_{0,a_1}\tau_{0,a_2}\tau_{0,a_3}}_0 = \delta_{a_1+a_2+a_3, r-2}.
\ee

\subsection{Generalized KdV hierarchies from $r$-spin curves}

Let us recall now the generalized Witten Conjecture for r-spin intersection numbers \cite{Witten2, Fab-Sha-Zvo}.
Introduce formal variables $t_{m,a}$ corresponding to $\tau_{m,k}$ ($m =0, 1,2, \dots$, $a=0, 1, \dots, r-2$),
consider
\bea
&& F_g(\bt) = \sum \corr{\tau_{m_1, a_1}  \dots  \tau_{m_n, a_n}}_g \cdot \frac{1}{n!} \prod_{i=1}^n t_{m_i, a_i}, \\
&& F(\bt; \lambda) = \sum_{g\geq 0} \lambda^{2g-2} F_g(\bt), \\
&& Z(\bt; \lambda) = \exp F(\bt;\lambda).
\eea
The following are the first few terms of $F_0$ and $F_1$:
\bea
&& F_0(\bt) = \frac{1}{3!} \sum_{a_1+a_2+a_3=r-2} t_{0,a_1}t_{0,a_2}t_{0,a_3} + \cdots, \\
&& F_1(\bt) = \frac{r-1}{24}t_{1,0} + \cdots.
\eea

The string equation and the dilaton equation can be reformulated as
the following two differential equations:
\bea
&& L_{-1} Z =0, \\
&& L_0 Z =0,
\eea
where $L_{-1}$ and $L_0$ are given by
\bea
&& L_{-1} = - \frac{\pd}{\pd t_{0,0}} + \sum_{k=1}^\infty \sum_{a=1}^2 t_{k,a} \frac{\pd}{\pd t_{k-1, a}}
+ \frac{1}{2\lambda^2} \sum_{a=0}^{r-2} t_{0,a} t_{0, r-2-a}, \\
&& L_{0} = - \frac{\pd}{\pd t_{1,0}} + \sum_{k=1}^\infty \sum_{a=1}^2 \frac{r k + a+1}{r+1} t_{k,a} \frac{\pd}{\pd t_{k, a}}
+ \frac{r-1}{24},
\eea
respectively.
The generalized Witten Conjecture can be stated as follows.
There is a pseudodifferential operator
\be
L = D^r + \sum_{i=0}^{r-2} u_i(\bt) D^i, \quad D = \frac{\sqrt{-1}}{\sqrt{r}} \frac{\pd}{\pd x}, \quad x = t_{0,0},
\ee
such that
\be
\frac{\pd^2 F}{\pd t_{0,0} \pd t_{n,a}} = - c_{n,a} \res(L^{n+\frac{a+1}{r}}),
\ee
where
\be
c_{n,a} = \frac{(-1)^n r^{n+1}}{(a+1)(a+1+r) \cdots (a+1 + nr)},
\ee
and
\be
\sqrt{-1} \frac{\pd L}{\pd t_{n, a}} = \frac{c_{n,a}}{\sqrt{r}} \cdot [(L^{n+(a+1)/r})_+, L].
\ee
Comparing with the notations in last section,
we set
\be
t_{n,a} = T_{nr+a+1} \cdot \frac{\sqrt{-r}}{c_{n,a}}
= (-1)^n \sqrt{-r} T_{nr+a+1} \cdot \prod_{j=0}^n (j+\frac{a+1}{r}).
\ee
Then in the new coordinates $\{T_1, \dots, T_{r-1}, T_{r+1}, \dots\}$,
$L$ satisfies
\be
\frac{\pd L}{\pd T_k} = [(L^{k/r})_+, L],
\ee
and
\be
\frac{\pd^2 F}{\pd T_1 \pd T_k} = \res (L^{k/r}).
\ee

\section{W-Constraints for  Witten's r-spin intersection numbers}

\subsection{String equation and dilaton equation in new coordinates}
The operators in string equation and the dilaton equation now become
\bea
&& L_{-1} = \sqrt{-r} \frac{\pd}{\pd T_1} -  \sum_{k=r+1}^\infty \frac{k}{r} T_k \frac{\pd}{\pd T_{k-r}}
- \frac{1}{2r \lambda^2} \sum_{b+c=r} bT_b \cdot cT_{c}, \\
&& L_{0} = - \frac{\sqrt{-r}}{1+\frac{1}{r}} \frac{\pd}{\pd T_{r+1}} +  \sum_{k=1}^\infty \frac{k}{r+1} T_{k} \frac{\pd}{\pd T_k}
+ \frac{r-1}{24},
\eea
We change them by multiplications of some constants and take now:
\bea
&& \tilde{L}_{-1} = \sum_{k=r+1}^\infty \frac{k}{r} \tilde{T}_k \frac{\pd}{\pd T_{k-r}}
+ \frac{1}{2r \lambda^2} \sum_{b+c=r} bT_b \cdot cT_{c}, \\
&& \tilde{L}_{0} = \sum_{k=1}^\infty \frac{k}{r} \tilde{T}_{k} \frac{\pd}{\pd T_k}
+ \frac{r^2-1}{24r},
\eea
where we have made the following dilaton shift:
\be
\tilde{T}_k = T_k - \delta_{k, r+1} \cdot \frac{\sqrt{-r}}{1+\frac{1}{r}}.
\ee

\subsection{W-Constraints for Witten's r-spin intersection numbers}
One can use the method Goeree \cite{Goeree} to derive the W-constraints for r-spin intersection numbers.
For an integer $r \geq 2$, let
\be
\alpha(z) = \sum_{n \in \bZ} \alpha_{\frac{n}{r}} z^{-\frac{n}{r}-1},
\ee
where
\be
 \alpha_m = \alpha_{-m} = 0, \quad m \in \bZ, \\
\ee
and for $m \geq 0$, $1 \leq j \leq r-1$,
\bea
&& \alpha_{m+\frac{j}{r}} = \lambda \frac{\pd}{\pd T_{rm +j}},  \\
&& \alpha_{-m-\frac{j}{r}} = \lambda^{-1} (rm+j) \tilde{T}_{rm+j},
\eea
It follows that one can rewrite $\alpha(z)$ as follows:
\be
\alpha(z) = \alpha_1(z) + \cdots + \alpha_{r-1}(z),
\ee
where
\be
\alpha_j(z) = \sum_{m \in \bZ} \alpha_{m+\frac{j}{r}} z^{-m-\frac{j}{r}-1},
\ee

Note
\bea
&& \tilde{L}_{-1} = \frac{1}{2r} \res_z :\alpha(z)\alpha(z):, \\
&& \tilde{L}_0 = \frac{1}{2r} \res_z z :\alpha(z)\alpha(z): + \frac{r^2-1}{24r}.
\eea

Consider the mode expansion of the following $r-1$ W-fields:
\bea
&& W^{(2)}(z) = \frac{1}{2!} :\alpha(z)^2: + \frac{r^2-1}{24 z^2}, \\
&& W^{(3)}(z) = \frac{1}{3!} :\alpha(z)^3:, \\
&& \cdots \cdots \cdots \cdots \cdots \cdots \\
&& W^{(r)}(z) = \frac{1}{p!} :\alpha(z)^r:.
\eea
I.e.,
write them as follows:
\be
W^{(k)}(z) = \sum_{m \in \bZ} \sum_{j=0}^{r-1} W^{(k)}_{m + \frac{j}{r}} z^{-m - \frac{j}{r}-k}.
\ee
Then the W-constraints satisfied by the partition function of Witten's r-spin intersection numbers are:
\be
W^{(k)}_m \tau = 0, \quad 2 \leq k \leq r, \quad m \geq -k+1.
\ee

Note
\be
W^{(k)}_m = \frac{1}{k!} \sum_{\substack{i_1, \dots, i_k \in 1/r \cdot \bZ \\
i_1 + \cdots + i_k = m }}
:\alpha_{i_1} \cdots \alpha_{i_k}:.
\ee

\section{Solution of the W-Constraints for R-Spin Intersection Numbers}

\subsection{Grading}
In Liu-Yang-Zhang \cite{Liu-Yang-Zhang},
the following grading for the r-spin case is used:
\be
\deg t_{n,a} = n + \frac{a+1}{r}.
\ee
For our purpose,
we define
\be \label{eqn:GradeTn}
\deg T_n = \frac{n}{r+1}.
\ee
The motivation for this definition is as follows.
First introduce some operators:
\be
\cO_{nr+a+1} = (-1)^n \sqrt{-r}  \cdot \prod_{j=0}^n (j+\frac{a+1}{r}) \cdot \tau_{n,a}.
\ee
They satisfy:
\be
t_{n,a} \tau_{n,a} = T_{nr+a+1} \cdot \cO_{nr+a+1}.
\ee
By the selection rule \eqref{eqn:SelRule} for
$\corr{ \tau_{m_1, k_1} \cdots \tau_{m_n, k_n} }_g$,
a correlator of the form $\corr{ \cO_{a_1} \cdots \cO_{a_n}}_g$ is nonvanishing  only if
\be
\frac{a_1}{r+1} + \cdots + \frac{a_n}{r+1} = 2g-2+n.
\ee
Hence one has
\be
F(\bT, \lambda) = \sum_{k=1}^\infty F^{(k)}(\bT, \lambda),
\ee
where
\be
F^{(k)}(\bT, \lambda) = \sum_{a_1+\cdots +a_n = (r+1)k} \lambda^{2g-2} \corr{\cO_{a_1} \cdots \cO_{a_n}}_g \prod_{j=1}^n T_{a_j}.
\ee
Furthermore,
one can write
\be
\tau(\bT)  = \sum_{k \geq 0} \tau^{(k)}(\bT),
\ee
where $\tau^{(k)}(\bT)$ has degree $k$,
and clearly $\tau^{(0)} = 1$.

Define the Euler operator
\be
E = \frac{1}{r+1} \sum_{n=1}^\infty nT_n \cdot \pd_{T_n}.
\ee
Then one has
\be
E \tau^{(k)} = k \cdot \tau^{(k)}.
\ee

\subsection{Gradings of the Virasoro operators}

Based on \eqref{eqn:GradeTn} we also define:
\be \label{eqn:DGradeTn}
\deg \frac{\pd}{\pd T_n} = - \frac{n}{r+1}.
\ee
Using these gradings
one can examine the gradings of the $W$-operators.
Let us first look at $W^{(2)}(z)$ first.
Its mode expansion is given by:
\ben
W^{(2)}(z) = \frac{1}{2} \sum_{i_1,i_2=1}^{r-1} \sum_{m_1, m_2\in \bZ}
:\alpha_{m_1+\frac{i_1}{r}} \alpha_{m_2+\frac{i_2}{r}}: z^{-(m_1+m_2+\frac{i_1+i_2}{r})-2} + \frac{r^2-1}{24z^2}.
\een
The relevant modes are:
\ben
&& W^{(2)}_{-1} = \half \sum_{m_1\in \bZ} \sum_{i_1=1}^{r-1} :\alpha_{m_1+\frac{i_1}{r}} \alpha_{-m_1-2+\frac{r-i_1}{r}}:, \\
&& W^{(2)}_0 = \half
\sum_{m_1\in \bZ} \sum_{i_1=1}^{r-1} :\alpha_{m_1+\frac{i_1}{r}} \alpha_{-m_1-1+\frac{r-i_1}{r}}:
+ \frac{r^2-1}{24}, \\
&& W^{(2)}_m = \frac{1}{2}
\sum_{m_1\in \bZ} \sum_{i_1=1}^{r-1} :\alpha_{m_1+\frac{i_1}{r}} \alpha_{m-m_1-1+\frac{r-i_1}{r}}:, \quad m > 0.
\een
Written explicitly as differential operators they are:
\bea
W^{(2)}_{-1} & = & - r\sqrt{-r} \frac{\pd}{\pd T_1} +  \sum_{k=r+1}^\infty kT_k \frac{\pd}{\pd T_{k-r}}
+ \frac{1}{2 \lambda^2} \sum_{b+c=r} bT_b \cdot cT_{c}, \\
W^{(2)}_{0} & = & - r\sqrt{-r} \frac{\pd}{\pd T_{r+1}} +  \sum_{k=1}^\infty k T_{k} \frac{\pd}{\pd T_k}
+ \frac{r^2-1}{24}, \\
W^{(2)}_m & = & - r \sqrt{-r} \frac{\pd}{\pd T_{(m+1)r+1}}
+ \sum_{k=1}^\infty kT_k \frac{\pd}{\pd T_{k+rm}}  \\
& + & \frac{\lambda^2}{2} \sum_{b+c=rm} \frac{\pd }{\pd T_b} \cdot \frac{\pd}{\pd T_c}. \nonumber
\eea
Now it is clear that one can write:
\be
W^{(2)}_m = W^{(2,0)}_m + W^{(2,1)}_m,
\ee
with
\be
\deg W_m^{(2,0)} = - \frac{rm}{r+1}, \quad \deg W^{(2,1)} =-1 -\frac{r m}{r+1}.
\ee

\subsection{Gradings of higher $W$-operators}

We will use the following notation:
$\sum'$ means a summation over some indices in $\frac{1}{r}\bZ$ non of which is $-\frac{r+1}{r}$ or integral.
We will also use operators
\be
\beta_{\frac{n}{r}} = \alpha_{\frac{n}{r}} + r \sqrt{-r} \delta_{n,-(r+1)} \lambda^{-1},
\ee
i.e.,
\bea
&& \beta_{m+\frac{j}{r}} = \lambda \frac{\pd}{\pd T_{rm +j}},  \\
&& \beta_{-m-\frac{j}{r}} = \lambda^{-1} (rm+j) T_{rm+j}.
\eea
It is clear that
\be
\deg \beta_i = \frac{ri}{r+1}.
\ee
Since one has
\ben
W^{(k)}_m
& = & \frac{1}{k!} \biggl( k(\alpha_{-(r+1)/r})^{k-1} \alpha_{m+(k-1)(r+1)/r} \\
& + & \binom{k}{2} (\alpha_{-(r+1)/r})^{k-2} \sum'_{i_1+i_2=m+(k-2)(r+1)/r}:\alpha_{i_1}\alpha_{i_2}: \\
& + & \cdots\cdots \cdots \cdots \\
& + &  \sum'_{i_1+\cdots + i_k =m}:\alpha_{i_1}\cdots \alpha_{i_k}: \biggr) + \delta_{k,2}\delta_{m,0} \frac{r^2-1}{24},
\een
after using $\alpha_{-(r+1)/r} = \beta_{-(r+1)/r} - r\sqrt{-1} \lambda^{-1}$ and $\alpha_i = \beta_i$ for $i \neq -(r+1)/r$,
one can rewrite $W^{(k)}_m$ in terms of the $\beta$-operators as follows:
\ben
W^{(k)}_m & = & \frac{1}{k!} \sum_{\substack{i_1, \dots, i_k \in 1/r \cdot \bZ \\ i_1 + \cdots + i_k = m}}
:\beta_{i_1} \cdots \beta_{i_k}: \\
& - & \frac{r\sqrt{-r}\lambda^{-1} }{(k-1)!}  \sum_{\substack{i_1, \dots, i_{k-1} \in 1/r \cdot \bZ \\ i_1 + \cdots + i_{k-1} = m+(r+1)/r}}
:\beta_{i_1} \cdots \beta_{i_{k-1}}: \\
& + & \frac{(r\sqrt{-r} \lambda^{-1})^2}{2!(k-2)!}  \sum_{\substack{i_1, \dots, i_{k-2} \in 1/r \cdot \bZ \\ i_1 + \cdots + i_{k-2} = m+2(r+1)/r}}
:\beta_{i_1} \cdots \beta_{i_{k-2}}: \\
& + & \cdots \cdots \cdots \cdots \cdots \\
& + &  \frac{(-r\sqrt{-r} \lambda^{-1} )^{k-2}}{(k-2)!2!}  \sum_{\substack{i_1, i_2 \in 1/r \cdot \bZ \\ i_1 +  i_2 = m+(k-2)(r+1)/r} }
:\beta_{i_1} \beta_{i_2}: \\
& + & \frac{(-r\sqrt{-r} \lambda^{-1} )^{k-1}}{(k-1)!1!}  \beta_{m+(k-1)(r+1)/r } + \delta_{k,2}\delta_{m,0} \frac{r^2-1}{24}.
\een
It follows that
\be
W^{(k)}_m = W^{(k,0)}_m + \cdots + W^{(k,k-1)},
\ee
where
\be
\begin{split}
W^{(k,j)}_m & =  \frac{(r\sqrt{-r} \lambda^{-1} )^{j}}{j!(k-j)!}  \sum_{\substack{i_1, \dots, i_{k-j} \in 1/r \cdot \bZ \\ i_1 + \cdots + i_{k-j} = m+j(r+1)/r}}
:\beta_{i_1} \cdots \beta_{i_{k-j}}:  \\
& + \delta_{k,2}\delta_{m,0}\delta_{j,0} \frac{r^2-1}{24}.
\end{split}
\ee
In particular,
\be
\begin{split}
W^{(k,k-1)}_m & =  \frac{(r\sqrt{-r} \lambda^{-1} )^{k-1}}{(k-1)!}  \beta_{m+(k-1)(r+1)/r} \\
& = \frac{(r\sqrt{-r} \lambda^{-1} )^{k-1}}{(k-1)!} \lambda \frac{\pd}{\pd T_{rm+(k-1)(r+1)}}.
\end{split}
\ee
Note
\be
\deg W_m^{(k,j)} = -\frac{rm}{r+1} - j.
\ee

\subsection{Solution of the W-constraints}

Now we use the grading introduced above to rewrite the $W$-constraint equations as follows:
\be
(W_m^{(k,0)} + \cdots + W^{(k,k-1)}_m) (\tau^{(0)} + \tau^{(1)} + \cdots ) = 0.
\ee
For each $j \geq 0$,
one then has:
\be
W_m^{(k,0)} \tau^{(j-k+1)} + W_m^{(k,1)} \tau^{(j-k+2)} + \cdots + W^{(k,k-1)}_m \tau^{(j)} = 0.
\ee
Or equivalently,
\be
\frac{\pd \tau^{(j)}}{\pd T_{rm+(k-1)(r+1)}} \\
=- \frac{(k-1)!\lambda^{k-2} }{(-r\sqrt{-r})^{k-1}}    \sum_{l=1}^{k-1} W_m^{(k,k-1-l)} \tau^{(j-l)}.
\ee
One can multiply both sides by $\frac{1}{r+1}(rm+(k-1)(r+1))T_{rm+(k-1)(r+1)}$ then take summation $\sum_{k=2}^r \sum_{m=-(k-1)}^\infty$ to get:
\ben
&& E\tau^{(j)} \\
&  =  &
- \sum_{k=2}^r \sum_{m=-(k-1)}^\infty \frac{(k-1)! \lambda^{k-2} }{(-r\sqrt{-r})^{k-1}}
\sum_{l=1}^{k-1} (\frac{rm}{r+1}+k-1) \\
&& \cdot T_{rm+(k-1)(r+1)} W_m^{(k,k-1-l)} \tau^{(j-l)} \\
& = & - \sum_{l=1}^{r-1} \sum_{k=l+1}^{r} \sum_{m=-(k-1)}^\infty \frac{(k-1)! \lambda^{k-2} }{(-r\sqrt{-r})^{k-1}} (\frac{rm}{r+1}+k-1) \\
&& \cdot T_{rm+(k-1)(r+1)} W_m^{(k,k-1-l)} \tau^{(j-l)}.
\een
One can rewrite it also as follows
\ben
E\tau^{(j)}
& = & - \frac{1}{r+1} \sum_{l=1}^{r-1} \sum_{k=l+1}^{r} \sum_{m=-(k-1)}^\infty \frac{(k-1)!\lambda^{k-1} }{(-r\sqrt{-r})^{k-1}} \\
&& \cdot \beta_{-(m+k-1 + (k-1)/r)} W_m^{(k,k-1-l)} \tau^{(j-l)}.
\een
After introducing the following operators for $l=1, \dots, r-1$:
\be
A_l
= - \frac{1}{r+1} \sum_{k=l+1}^{r} \sum_{m=0}^\infty \frac{(k-1)!\lambda^{k-1}}{(-r\sqrt{-r})^{k-1}} \beta_{-(m+(k-1)/r)} W_{m-k+1}^{(k,k-1-l)}
\ee
one gets

\begin{thm}
The partition function of $r$-spin intersection numbers can be computed recursively as follows:
\be
j \tau^{(j)} =  \sum_{l=1}^{r-1} A_l \tau^{(j-l)}.
\ee
\end{thm}

Together with the initial value $\tau^{(0)} = 1$,
this then provides a solution of the W-constraints
for the r-spin intersection numbers.
We conjecture that
\be
[A_i, A_j] = 0
\ee
for $i,j=1, \dots, r-1$.
If this is true,
then we have
\be
\tau = \exp \biggl( \sum_{j=1}^{r-1} \frac{1}{j} A_j \biggr) 1.
\ee

\section{Examples}

\subsection{The $r=2$ case}
In this case,
\be
\begin{split}
A_1 & = \frac{\lambda}{6\sqrt{-2}}
\biggl(\sum_{\substack{a, b \in \half \bZ_+\\a+b \geq \frac{3}{2} }} \beta_{-a}\beta_{-b} \beta_{a+b-\frac{3}{2}}
+ \frac{1}{2} \sum_{a,b \in \half\bZ_+} \beta_{-a-b-\frac{3}{2}} \beta_{a}\beta_{b} \\
& + \frac{1}{2} \beta_{-\frac{1}{2}}^3 + \frac{1}{8} \beta_{-\frac{3}{2}}  \biggr).
\end{split}
\ee
Our result gives:
\be
\tau = e^{A_1} 1.
\ee
This matches with Alexandrov \cite{Alexandrov}.

\subsection{The $r=3$ case}

We use a Maple program by Hao Xu to compute the 3-spin intersection numbers.
For $g=0$ and $n=3$,
\ben
&& \corr{\tau_{0,0}\tau_{0, 0} \tau_{0,1}}_0 = 1;
\een
for $g=0$ and $n=4$,
\ben
&& \corr{\tau_{0,1}^4}_0 = \frac{1}{3}, \;\; \corr{\tau_{1,1}\tau_{0,0}^3}_0 = 1, \;\; \corr{\tau_{1,0}\tau_{0,1}\tau_{0,0}^2}_0 =1;
\een
for $g=0$ and $n=5$,
\ben
&& \corr{\tau_{2,1}\tau_{0,0}^4}_0 = 1, \;\; \corr{\tau_{2,0}\tau_{0,1}\tau_{0,0}^3}_0 = 1, \\
&& \corr{\tau_{1,1}\tau_{1,0}\tau_{0,0}^3}_0 =2, \;\;
 \corr{\tau_{1,0}^2\tau_{0,1}\tau_{0,0}^2}_0 = 2;
\een
hence one has
\ben
F_0 & = & \frac{1}{2} t_{0,0}^2 t_{0,1} + \biggl(\frac{1}{72} t_{0,1}^4 + \frac{1}{6} t_{1,1}t_{0,0}^3
+ \frac{1}{2} t_{1,0}t_{0,1}t_{0,0}^2 \biggr) \\
& + & \biggl( \frac{1}{24} t_{2,1}t_{0,0}^4 +  \frac{1}{6} t_{2,0}t_{0,1}t_{0,0}^3
+ \frac{1}{3} t_{1,1}t_{0,1}t_{0,0}^3 + \frac{1}{2} t_{1,0}^2t_{0,1}t_{0,0}^2 \biggr) + \cdots
\een
For $g=1$ and $n=1$,
\ben
&& \corr{\tau_{1,0}}_1 = \frac{1}{12};
\een
for $g=1$ and $n=2$,
\ben
&& \corr{\tau_{2,0} \tau_{0,0} }_1 = \frac{1}{12}, \quad \corr{ \tau_{1,0}\tau_{1,0} }_1 = \frac{1}{12};
\een
for $g=1$ and $n=3$,
\ben
&& \corr{\tau_{3,0} \tau_{0,0}^2 }_1 = \frac{1}{12}, \quad \corr{ \tau_{2,0}\tau_{1,0}\tau_{0,0} }_1 = \frac{1}{6},
\quad \corr{ \tau_{1,0}^3}_1 = \frac{1}{6}; \\
&& \corr{\tau_{2,1} \tau_{0,1}^2 }_1 = \frac{1}{36}, \quad \corr{ \tau_{1,1}^2\tau_{0,1} }_1 = \frac{1}{36};
\een
hence
\ben
F_1 & = & \frac{1}{12} t_{1,0}
+ \biggl(\frac{1}{12} t_{2,0} t_{0,0} + \frac{1}{24} t_{1,0}^2 \biggr) \\
& + & \biggl( \frac{1}{24} t_{3,0} t_{0,0}^2 + \frac{1}{6} t_{2,0}t_{1,0}t_{0,0}
+ \frac{1}{36} t_{1,0}^3
+ \frac{1}{72} t_{2,1} t_{0,1}^2 + \frac{1}{72} t_{1,1}^2t_{0,1} \biggr) + \cdots.
\een
For $g=2$,
there is no nonvanishing intersection numbers for $n=1$;
for $g=2$ and $n=2$,
\ben
\corr{\tau_{4,1} \tau_{0,1}}_2 = \frac{1}{864}, \quad
\corr{\tau_{3,1} \tau_{1,1}}_2 = \frac{11}{4320}, \quad
\corr{\tau_{2,1}^2}_2 = \frac{17}{4320},
\een
and so
\ben
F_2 = \biggl( \frac{1}{864} t_{4,1}t_{0,1} + \frac{11}{4320} t_{3,1}t_{1,1} + \frac{17}{8640} t_{2,1}^2 \biggr) + \cdots.
\een

After the change of coordinates:
\ben
t_{n,a} = (-1)^n \sqrt{-3} \prod_{j=0}^n (j+\frac{a+1}{3}) \cdot T_{3n+a+1},
\een
one then has
\ben
F_0 & = & -\frac{\sqrt{-3}}{9} T_2T_1^2 + \biggl(\frac{2}{81}T_2^4-\frac{5}{81}T_5T_1^3-\frac{4}{27}T_4T_2T_1^2 \biggr) \\
& + & \sqrt{-3} \biggl(\frac{2}{243}T_8 T_1^4 + \frac{28}{729}T_7 T_2T_1^3-\frac{20}{243}T_5T_2T_1^3
+\frac{16}{243}T_4^2T_2T_1^2 \biggr) + \cdots, \\
F_1 & = &  -\frac{\sqrt{-3}}{27} T_4- \biggl( \frac{7}{81} T_7T_1+\frac{2}{81}T_4^2 \biggr) \\
& + & \sqrt{-3} \biggl( \frac{56}{729} T_7T_4T_1+\frac{16}{2187} T_4^3
- \frac{40}{729} T_8T_2^2-\frac{25}{729} T_5^2T_2 \biggr) + \cdots, \\
F_2 & = & -\frac{770}{6561}T_{14}T_2-\frac{605}{6561} T_{11}T_5-\frac{340}{6561} T_8^2 + \cdots
\een
One can see that
\ben
F^{(1)} & = & -\frac{\sqrt{-3}}{9} T_2T_1^2 \lambda^{-2}-\frac{\sqrt{-3}}{27} T_4, \\
F^{(2)} & = & \biggl(\frac{2}{81}T_2^4-\frac{5}{81}T_5T_1^3-\frac{4}{27}T_4T_2T_1^2 \biggr) \lambda^{-2}
- \biggl( \frac{7}{81} T_7T_1+\frac{2}{81}T_4^2 \biggr),  \\
F^{(3)} & = & \sqrt{-3} \biggl(\frac{2}{243}T_8 T_1^4 + \frac{28}{729}T_7 T_2T_1^3-\frac{20}{243}T_5T_2T_1^3
+\frac{16}{243}T_4^2T_2T_1^2 \biggr) \lambda^{-2} \\
& + & \sqrt{-3} \biggl( \frac{56}{729} T_7T_4T_1+\frac{16}{2187} T_4^3
- \frac{40}{729} T_8T_2^2-\frac{25}{729} T_5^2T_2 \biggr),
\een
It follows  that
\ben
&& \tau^{(1)} = -\frac{\sqrt{-3}}{9} T_2T_1^2 \lambda^{-2}-\frac{\sqrt{-3}}{27} T_4, \\
&& \tau^{(2)} =  -\frac{\lambda^{-4}}{54}T_2^2T_1^4
- \lambda^{-2} \biggl( \frac{13}{81} T_4T_2T_1^2 - \frac{2}{81}T_2^4 + \frac{5}{81} T_5T_1^3\biggr)
-\frac{13}{486}T_4^2-\frac{7}{81}T_7T_1, \\
&& \tau^{(3)} = \sqrt{-3} \biggl(
\frac{\lambda^{-6}}{1458}T_2^3T_1^6
+ \lambda^{-4} \biggl( \frac{25}{1458} T_2^2T_1^4T_4/x^4-\frac{2}{729} T_2^5T_1^2+\frac{5}{729} T_2T_1^5T_5 \biggr) \\
&& + \lambda^{-2} \biggl( \frac{2}{243} T_8T_1^4- \frac{2}{2187} T_4T_2^4+\frac{325}{4374}T_4^2T_2T_1^2
+ \frac{35}{729} T_7T_2T_1^3-\frac{20}{243} T_5T_2T_1^3+\frac{5}{2187} T_4T_5T_1^3 \biggr) \\
&& -\frac{40}{729} T_8T_2^2
- \frac{25}{729} T_5^2T_2+\frac{325}{39366} T_4^3
+ \frac{175}{2187}T_7T_4T_1.
\een

On the other hand, in this case
\ben
&& A_1
= \frac{\lambda}{12\sqrt{-3}} \sum_{m=0}^\infty \beta_{-(m+1/3)} W^{(2,0)}_{m-1}
+ \frac{\lambda^2}{54} \sum_{m=0}^\infty \beta_{-(m+2/3)} W^{(3,1)}_{m-2}, \\
&& A_2
= \frac{\lambda^2}{54} \sum_{m=0}^\infty  \beta_{-(m+\frac{2}{3})} W_{m-2}^{(3,0)},
\een
where the components of the $W$-operators are given by:
\ben
W^{(2,0)}_m  & = & \frac{1}{2}  \sum_{\substack{i_1,   i_2 \in 1/3 \cdot \bZ \\ i_1 + i_2 = m}}
:\beta_{i_1} \beta_{i_2}: + \frac{\delta_{m,0}}{3}, \\
W^{(3,0)}_m  & = & \frac{1}{6}  \sum_{\substack{i_1, \dots, i_3 \in 1/3 \cdot \bZ \\ i_1 + \cdots + i_{3} = m}}
:\beta_{i_1} \beta_{i_2} \beta_{i_3}:, \\
W^{(3,1)}_m  & = & \frac{3\sqrt{-3} \lambda^{-1}}{2}  \sum_{\substack{i_1,   i_2 \in 1/3 \cdot \bZ \\ i_1 + i_2 = m+4/3}}
:\beta_{i_1} \beta_{i_2}:.
\een
It follows that
\ben
A_1
& = & \frac{\lambda}{24\sqrt{-3}} \sum_{m=0}^\infty \beta_{-(m+1/3)} \sum_{\substack{i_1,   i_2 \in 1/3 \cdot \bZ \\ i_1 + i_2 = m-1}}
:\beta_{i_1} \beta_{i_2}:
+ \frac{\lambda}{36\sqrt{-3}} \beta_{-4/3} \\
& + & \frac{\lambda}{12\sqrt{-3}} \sum_{m=0}^\infty \beta_{-(m+2/3)} \sum_{\substack{i_1,   i_2 \in 1/3 \cdot \bZ \\ i_1 + i_2 = m-2/3}}
:\beta_{i_1} \beta_{i_2}:.
\een

\ben
A_2
& = & \frac{\lambda^2}{324} \sum_{m=0}^\infty  \beta_{-(m+\frac{2}{3})}
\sum_{\substack{i_1, \dots, i_3 \in 1/3 \cdot \bZ \\ i_1 + \cdots + i_{3} = m-2}}
:\beta_{i_1} \beta_{i_2} \beta_{i_3}:
\een
They can be written more explicitly as follows:
\ben
&& A_1
= \frac{\lambda}{36\sqrt{-3}} \beta_{-4/3} + \frac{\lambda}{12\sqrt{-3}} \biggl(
2 \beta_{-2/3} \beta_{-1/3}\beta_{-1/3} \\
& + &  \sum_{m=0}^\infty \sum_{b+c= m+ \frac{5}{3} } \beta_{-b}\beta_{-c} \beta_{m+\frac{1}{3}}
+ \frac{3}{2} \sum_{m=0}^\infty \sum_{b+c= m+2} \beta_{-b}\beta_{-c} \beta_{m+\frac{2}{3}} \\
& + & \half \sum_{m=0}^\infty \beta_{-(m+\frac{7}{3})} \sum_{b+c=m+1} \beta_b\beta_c
+  \sum_{m=0}^\infty \beta_{-(m+\frac{8}{3})} \sum_{b+c=m+\frac{4}{3}} \beta_b\beta_c \biggr),
\een
and
\ben
A_2
& = & \frac{\lambda}{324} \biggl\{
\beta_{-2/3}^4 + 3 \beta_{-4/3}\beta_{-\frac{2}{3}} \beta_{-1/3}^2  + \beta_{-5/3}\beta_{-1/3}^3  \\
&& + 9 \beta_{-\frac{5}{3}} \beta_{-\frac{2}{3}}^2 \beta_{\frac{1}{3}} \\
&& + (3\beta_{-\frac{8}{3}} \beta_{-\frac{1}{3}}^2 + 6 \beta_{-\frac{7}{3}} \beta_{-\frac{2}{3}} \beta_{-\frac{1}{3}}
+ 6 \beta_{-\frac{5}{3}} \beta_{-\frac{4}{3}} \beta_{-\frac{1}{3}} +3\beta_{-\frac{4}{3}}^2 \beta_{-\frac{2}{3}} ) \beta_{\frac{2}{3}} \\
& + & (9\beta^2_{-\frac{5}{3}}\beta_{-\frac{2}{3}} + 6 \beta_{-\frac{8}{3}}\beta_{-\frac{2}{3}}^2) \beta_{\frac{4}{3}} + \cdots \biggr\}
\een
It follows that:
\ben
A_11 & = & \frac{\lambda}{36\sqrt{-3}} \beta_{-4/3} + \frac{\lambda}{6\sqrt{-3}}   2 \beta_{-2/3} \beta_{-1/3}\beta_{-1/3} \\
& = & \frac{1}{9\sqrt{-3}}T_4 + \frac{1}{3\sqrt{-3}\lambda^2} T_2T_1^2,
\een

\ben
A_1^21 & = & \biggl(\frac{1}{9\sqrt{-3}}T_4 + \frac{1}{3\sqrt{-3}\lambda^2} T_2T_1^2\biggr)^2 \\
& + & \frac{\lambda}{12\sqrt{-3}} \biggl(\frac{1}{9 \sqrt{-3}} \sum_{b+c=1+ \frac{5}{3}} \beta_{-b}\beta_{-c} \lambda
+ \frac{1}{3 \sqrt{-3} \lambda^2} \sum_{b+c=\frac{5}{3}} \beta_{-b}\beta_{-c} 2\lambda T_2T_1 \\
&& + \frac{3}{2} \frac{1}{3 \sqrt{-3}\lambda^2}  \sum_{b+c=2} \beta_{-b}\beta_{-c} T_1^2 \biggr) \\
& + & \frac{\lambda}{12\sqrt{-3}} \cdot \frac{1}{2} \beta_{-\frac{7}{3}} 2\lambda^2 \pd_{T_1}\pd_{T_2} (\frac{1}{3\sqrt{-3}\lambda^2} T_2T_1^2) \\
& = & - \frac{13}{243}T_4^2-\frac{14}{81}T_7T_1
-\frac{5}{36}T_5T_1^3\lambda^{-2}-\frac{32}{81}T_4T_2T_1^2 \lambda^{-2}-\frac{1}{27}T_2^2T_1^4\lambda^{-4}.
\een

\ben
A_2 1
& = & \frac{\lambda^2}{324} \biggl\{
\beta_{-2/3}^4 + 3 \beta_{-4/3}\beta_{-\frac{2}{3}} \beta_{-1/3}^2  + \beta_{-5/3}\beta_{-1/3}^3 \biggr\} 1 \\
& = & \frac{\lambda^{-2}}{324} (16T_2^4 + 24 T_4T_2T_1^2 + 5 T_5T_1^3).
\een

It can be checked that
\ben
&& \tau^{(1)} = A_1 1, \\
&& \tau^{(2)} = \half (A_1^21 + A_2 1).
\een

\vspace{.2in}
{\em Acknowledgements}.
The author thanks Professor Hao Xu for help on computing r-spin numbers.
He also thanks the authors of \cite{Liu-Yang-Zhang} for explaining their work.
This research is partially supported by NSFC grant 1171174.

\end{document}